\documentstyle[11pt,newpasp_spekkens,twoside,epsf]{article}
\def\edcomment#1{\iffalse\marginpar{\raggedright\sl#1\/}\else\relax\fi}
\reversemarginpar

\begin{document}
\title{Inner Halo Shapes of Dwarf Galaxies: Resolving the Cusp/Core Problem}
 \author{Kristine Spekkens \& Riccardo Giovanelli}
\affil{Cornell University, Space Sciences Building, Ithaca, NY 14853, USA}

\begin{abstract}
 We derive inner dark matter halo density profile slopes for a sample of 200 dwarf galaxies by inverting rotation curves obtained from high-quality, long-slit optical spectra. Using simulations to assess the impact of long-slit observing and data processing errors on our measurements, we conclude that our observations are consistent with the cuspy halos predicted by the CDM paradigm.
\end{abstract}

\section{Introduction}
\label{intro}
While the hierarchical CDM paradigm is very successful at reproducing observations on scales greater than a few Mpc, the agreement between between CDM predictions and galaxy halo properties is not as certain. In particular, measured inner halo density profile slopes of dwarf galaxies inferred from long-slit optical spectra tend to be shallower than the cusps obtained from CDM simulations of halo assembly ({\it e.g.} de Blok, Bosma, \& McGaugh 2003, hereafter dBBM; Swaters et al. 2003, hereafter SMBB). The implications of this cusp/core problem in light of observational uncertainties remains unclear: while some authors advocate a genuine conflict between theory and observations (dBBM), others claim consistency between the data and cuspy CDM halos (SMBB). In this paper, we obtain inner density profile slopes for a large sample of dwarf galaxies and investigate the impact of observational and data processing errors on our result with detailed simulations. 

\section{Sample Selection and Data Analysis}
\label{sample}
 We select 200 low-mass galaxies from the SFI++ catalog, a 4800-object Tully-Fisher database maintained at Cornell University (Catinella et al. 2003). Galaxies in the sample have $V_{rot}<130\,\rm{km}\,\rm{s}^{-1}$, archived rotation curves (RCs) from high-quality H$\alpha$ spectroscopy and no evidence for a bulge, bar, or other baryonic distortions as determined by our accurate I-band photometry. 

 For each rotation curve, the density $\rho(r_i)$ is derived at each RC point $r_i$ assuming a minimal disk and spherical symmetry: $\rho(r_i)\propto 2(V_i/r_i)(\textrm{d}V/\textrm{d}r)_i + V_i^2/r_i^2$. The rotation velocity $V_i$ and RC derivative $(\textrm{d}V/\textrm{d}r)_i$ are obtained from a best-fit smooth curve to the RC points. For $\rho(r)\propto r^{-\alpha}$ at small r, we measure the inner slope $\alpha_m$ from a linear fit to the inner 2-3 points of log($\rho(r)$). Fig.~1 shows our results (solid lines): $\alpha_m$ is generally smaller than the intrinsic slope $\alpha_{int}\sim\,1$ predicted by CDM simulations. 

\section{Dwarf Population Simulations}
\label{sim}

 To investigate the consistency of our findings with CDM predictions, we simulate long-slit observations of 200 dwarf galaxies with intrinsic cusps. We embed infinitely thin, uniform H$\alpha$ disks inside NFW halos ($\alpha_{int}=1$; {\it e.g.} Navarro, Frenk \& White 1996).  The primary observational parameters for each system are random deviates of corresponding distributions derived for the sample. The simulated galaxies are then ``observed'' (with some error, chosen to match RC folding outputs from the sample) in typical conditions. The resulting RCs are calibrated and $\alpha_m$ is measured using the algorithm described in \S2. 

Fig.~1 shows the distribution of $\alpha_m$ obtained for the simulated dwarf population (dashed lines) superimposed on the sample distribution (solid lines); on average, the $\alpha_m$ inferred for the simulated halos are shallower than the intrinsic slope $\alpha_{int}=1$, and resemble those obtained for the galaxy sample.  Quantitatively, a Kolmogorov-Smirnov test returns a ``P-value'' of 0.1, indicating no statistically significant difference between the two distributions. We conclude that these long-slit spectroscopic observations are consistent with cuspy halos predicted by the CDM paradigm without recourse to halo triaxiality, modifications of Newtonian dynamics or other exotic phenomena.

\begin{figure}
\plotfiddle{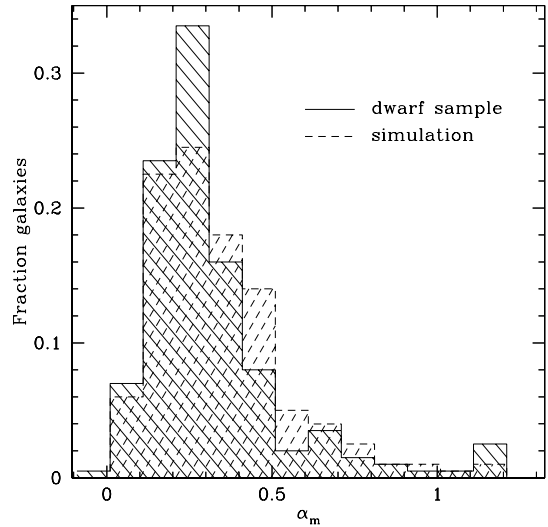}{180pt}{0}{37}{37}{-115}{-65}
\caption{Inner slopes $\alpha_m$ for the dwarf sample (solid lines) and simulated galaxies (dashed lines). See text for details.}
\end{figure}

\end{document}